\documentclass[prd,aps,floats]{revtex4}
\usepackage{amsmath}
\usepackage{amssymb}

\numberwithin{equation}{section}

\usepackage{epsfig}

\def\beq{\begin{equation}}
\def\eeq{\end{equation}}
\def\ber{\begin{eqnarray}}
\def\eer{\end{eqnarray}}

\def\om{\Omega_{0m}}
\def\atridot{\stackrel{...}{a}}
\def \lleq {\lower0.9ex\hbox{ $\buildrel < \over \sim$} ~}
\def \ggeq {\lower0.9ex\hbox{ $\buildrel > \over \sim$} ~}
\def\apj{{Astroph.\@ J.\ }}
\def\mn{{Mon.\@ Not.\@ Roy.\@ Ast.\@ Soc.\ }}

\def\prl{{Phys.\@ Rev.\@ Lett.\ }}
\def\prd{{Phys.\@ Rev.\@ D\ }}

\def\plb {{Phys.\@ Lett.\@ B\ }}
\def \jetpl {JETP Lett.\ }
\def\etal{{\it et al.}}

\begin{document}

\title{Two new diagnostics of dark energy}

\author{Varun Sahni$^a$, Arman Shafieloo$^a$ and Alexei A. Starobinsky$^b$}
\affiliation{$^a$ Inter-University Centre for Astronomy and Astrophysics,
Post Bag 4, Ganeshkhind, Pune 411~007, India}
\affiliation{$^b$Landau Institute for Theoretical Physics, Kosygina 2,
Moscow 119334, Russia
}

\thispagestyle{empty}

%\maketitle
\sloppy

\begin{abstract}
We introduce two new diagnostics of dark energy (DE). The first, $Om$,
is a combination of the Hubble parameter and the cosmological redshift
and provides a {\em null test} of dark energy being a cosmological constant
$\Lambda$. Namely, if the value of $Om(z)$ is the same at different
redshifts, then DE $\equiv \Lambda$, {\em exactly}.
%The $Om$ diagnostic -- a combination of the Hubble parameter and the
%cosmological redshift -- can help in distinguishing between the cosmological
%constant ($w = -1$) and dynamical dark energy ($w \neq -1$).
The slope of $Om(z)$ can differentiate between different models of
dark energy even if the value of the matter density is not
accurately known. For DE with an unevolving equation of state,
a positive slope of $Om(z)$ is suggestive of Phantom ($w < -1$) while a
negative slope indicates Quintessence ($w > -1$).
The second diagnostic -- {\em acceleration probe} $\bar q$ -- is the mean value
of the deceleration parameter over a small redshift range. It can be used to
determine the cosmological redshift at which the universe began to accelerate,
again without reference to the current value of the matter density.
We apply the $Om$ and $\bar q$ diagnostics to the Union
data set of type Ia supernovae combined with recent data from the cosmic microwave background
(WMAP5) and 
baryon acoustic oscillations.

\end{abstract}

\maketitle

%\pacs{PACS number(s): 04.50.+h, 98.80.Hw}

\bigskip

\section{Introduction}
\label{sec:intro}
The nature of dark energy (DE) is one of the most intriguing questions facing
physics. The fact that DE provides the main contribution to the energy
budget of the universe today while remaining subdominant during previous epochs, provides a
challenge to model builders attempting to understand the nature of this seemingly
all-pervasive ether-like substance.

Theoretical models for DE include the famous cosmological constant,
$\Lambda$, suggested by Einstein in 1917 \cite{einstein} and shown
to be related to the vacuum energy $\langle T_{ik} \rangle_{\rm vac}
\propto \Lambda g_{ik}$ several decades later \cite{zel68,wein89}.
Indeed, the cosmological constant appears to occupy a privileged
position amongst other DE models by virtue of the fact that its
equation of state $w=-1$ is Lorentz invariant and so appears the
same to any inertial observer. However, within the context of
cosmology, an explanation of DE in terms of $\Lambda$ faces one
drawback, namely, in order for the universe to accelerate today the
ratio of the energy density in the cosmological constant to that in
radiation must have been very small at early times, for instance
$\rho_\Lambda \simeq 3\times 10^{-58}\rho_{\rm EW}$ at the time of
the electroweak phase transition. Although vacuum energy may
conceivably be associated with small numbers such as the neutrino
mass ($\rho_\Lambda \sim m_\nu^4$) or even the fine structure
constant \cite{S99}, a firm theoretical prediction for the value of
$\Lambda$ is currently lacking, allowing room for alternatives
including models in which both the DE density and its equation of
state (EOS) evolve with time. Alternatives to the cosmological
constant include scalar field models called quintessence which have
$w > -1$, as well as more exotic `phantom' models with $w < -1$.
%Tracker fields belonging to the quintessence
%variety have attracted much recent attention due to the fact that
%such fields follow a common evolutionary path from a wide range of
%initial conditions. This is sometimes seen as providing a remedy to the initial
%conditions problem faced by the unevolving $\Lambda$-term.

Although most recent studies show that a cosmological constant + cold dark
matter (LCDM) is in excellent agreement with observational data, dynamical dark
energy can explain the data, too
\cite{observations1,snls,HST,essence,union,wmap03,wmap05,metamorphosis,observations2,alam07,arman2,DE_review,ss06}.
%In fact, given the enormous variety of DE models which have been suggested,
%the need of the hour appears to be a diagnostic capable of differentiating
%the cosmological constant from `something else'.
Indeed, the enormous variety of DE models suggested in the literature (see
\cite{DE_review} for reviews) has been partially responsible for the burgeoning
industry of model independent techniques aimed at reconstructing the properties
of dark energy directly from observations \cite{ss06}. It is well known that
model independent methods must be wary of several pitfalls which can subvert
their efficacy. These relate to priors which are sometimes assumed about
fundamental cosmological quantities such as the EOS and the matter density. As
first pointed out in \cite{maor02}, an incorrect prior for the EOS
%(such as demanding that it be unevolving) can lead to
can lead to gross misrepresentations of reality. The same applies to the value
of the matter density. Indeed, as we shall demonstrate later in this paper, an
incorrect assumption about the value of $\om$ can lead to dramatically
incorrect conclusions being drawn about the nature of dark energy.
%The present paper belongs to this category. We introduce
Clearly the need of the hour, then, is a diagnostic which is able to
differentiate LCDM from `something else' with as few priors as possible being
set on other cosmological parameters.

In this paper we introduce a new diagnostic, $Om$, which is
constructed from the Hubble parameter $H\equiv \dot a/a$ determined
directly from observational data and provides a {\em null test} of
the LCDM hypothesis. Here $a(t)$ is the scale factor of a
Friedmann-Robertson-Walker (FRW) cosmology. We show that $Om$ is
able to distinguish dynamical DE from the cosmological constant in a
robust manner {\em both with and without reference} to the value of
the matter density, which can be a significant source of uncertainty
for cosmological reconstruction. The $Om$ diagnostic is in many
respects the logical companion to the statefinder $r = \atridot/a
H^3$ \cite{statefinder} (otherwise dubbed jerk $j$, see e.g.
\cite{V04,DG08} as well as the earlier paper \cite{CN98}). We remind
the reader that $r = 1$ for LCDM while $ r \neq 1$ for evolving DE
models. Hence $r(z_1) - r(z_2)$ provides a {\em null test} for the
cosmological constant. Similarly, the unevolving nature of $Om(z)$
in LCDM furnishes $Om(z_1) - Om(z_2)$ as a {\em null test} for the
cosmological constant. 
(For null tests based on gravitational clustering see \cite{CN07,np08}.)
Like the statefinder, $Om$ depends only upon
the expansion history of our Universe. However, while the
statefinder $r$ involves the third derivative of the expansion
factor $a(t)$, $Om$ depends upon its {\em first} derivative only.
Therefore, as
we demonstrate in this paper, $Om$ is much easier to reconstruct
from observations
\footnote{Note that the $Om$ diagnostic does not
use any information about the evolution of inhomogeneities 
in an FRW background, usually 
represented by the growth factor $\delta(z)\equiv
(\delta\rho/\rho)_m$. Therefore our proposal for a null test of
LCDM is totally unrelated to earlier null tests 
\cite{CN07,np08} (based on the analytical expression for
$\Omega_m$ in terms of $\delta(z)$ obtained in \cite{st98}), and to 
the test of physical nature of dark energy (equivalently, that
$G_{eff}$ is a constant) suggested in \cite{ss06}.} .

It is clear that a determination of $H(z)$ from any
single test may suffer from systematic uncertainties. That is why,
in order to obtain $Om$ with sufficient accuracy, it is necessary to combine
information about $H(z)$ obtained from different independent tests,
such as the supernovae luminosity distance, the scale of baryon
acoustic oscillations (BAO) in the matter power spectrum as a
function of $z$, the acoustic scale in the angular power spectrum of
the cosmic microwave background (CMB) temperature fluctuations, etc.

The second diagnostic -- {\em Acceleration probe} $\bar q$ -- is constructed
out of the Hubble parameter and the lookback time. Like $Om$ it does not depend
upon the current value of the matter density. We apply $\bar q$ to current data
and show that it provides an independent test of the present acceleration of
the Universe.

The plan of the rest of the paper is as follows. In Sec. 2
the $Om$ diagnostic is introduced for a flat as well as a spatially curved
FRW background, and determined using SNLS and Union supernovae data. In
Sec. 3 the $\bar q$ diagnostic is introduced. In Sec. 4 both these
diagnostics are determined from a combination of existing data sets including
type Ia supernovae, BAO and WMAP5 CMB data. Sec. 5 contains our conclusions.

%\section{Influence of the matter density on properties of dark energy}
\section{The Om diagnostic -- a null test of LCDM}
\subsection{Influence of $\om$ on properties of dark energy}

\begin{figure*}[ht]
\centerline{ \psfig{figure=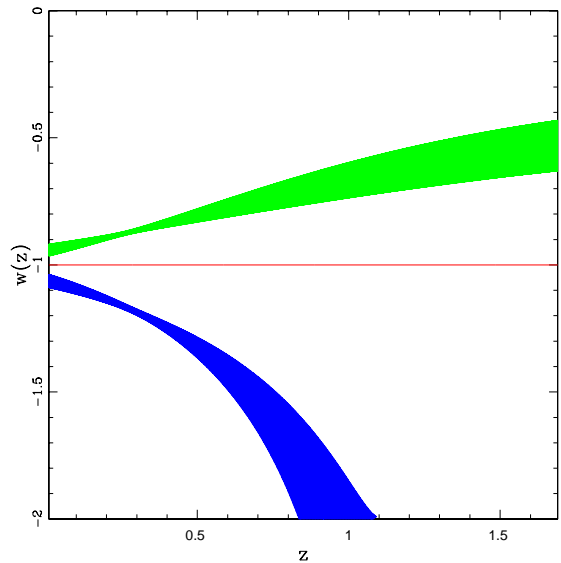,width=0.45\textwidth,angle=0} }
\bigskip
\caption{\small
The equation of state of a fiducial LCDM model ($w=-1, \om^{true}=0.27$)
is reconstructed
using an incorrect value of the matter density.
For $\om^{erroneous} = 0.22$ the resulting EOS shows quintessence-like behavior
and its $1-\sigma$ contour is shown in green.
In the opposite case, when $\om^{erroneous} = 0.32$, the EOS is
phantom-like and its $1-\sigma$ contour is shown in blue.
Note that in both cases the true fiducial model (red) is excluded in the reconstruction.
(The parametric reconstruction scheme suggested
 in \cite{statefinder} was applied to SNAP-quality data
to construct this figure.)
%The change from using other ansatz's is expected to be marginal.
}
\label{fig:eos}
\end{figure*}

Given many alternative models of dark energy it is useful
to try and understand the properties of DE in a model independent manner.
An important model independent quantity is the expansion history, $H(z)$,
whose value can be reconstructed from observations of the luminosity distance, $D_L$,
via a single differentiation
\cite{st98,HT99,chiba99,saini00,chiba00}
\beq\label{eq:H}
H(z)=\left[{d\over dz}\left({D_L(z)\over 1+z}\right)\right]^{-1}~.
\eeq
The equation of state, $w(z)$, of DE is more cumbersome to reconstruct since it
involves two derivatives of $D_L(z)$ and is therefore a noisier quantity
than $H(z)$. An additional source of uncertainty relating to $w(z)$
is caused by the fact that the value of the matter density, $\Omega_{0m}$ enters
into the determination of $w(z)$ explicitly, through the expression
\beq\label{eq:state}
w(x) =
\frac{(2 x /3) \ d \ {\rm ln}H \ / \ dx - 1}{1 \ - \ (H_0/H)^2
\Omega_{0m} \ x^3}\,\,.
\eeq
%Since $\om$ sits in the denominator of (\ref{eq:state}),
Clearly an uncertainty in $\om$ propagates into the EOS of dark energy
even if $H(z)$ has been reconstructed quite accurately.
This fact has been emphasized in several papers \cite{maor02,alam07,ss06,arman,kunz}
and is illustrated in figure \ref{fig:eos},
which shows how an erroneous estimate of $\om$
adversely affects the reconstructed EOS by making a LCDM model appear as if
it were quintessence (if $\om^{erroneous} < \om^{true}$) or phantom
(if $\om^{erroneous} > \om^{true}$).

The influence of dark matter on dark energy persists if a parametric ansatz such as CPL \cite{CPL}
\beq
w(z) = w_0 + w_1\frac{z}{1+z},
\label{eq:cpl}
\eeq
is employed in the determination of
\ber
H^2(z) &=& H_0^2 \lbrack \om (1+z)^3 + \Omega_{\rm DE}\rbrack,\nonumber\\
\nonumber\\
\Omega_{\rm DE} &=& (1-\om) \exp{\left\lbrace 3 \int_0^{z}
\frac{1+w(z')}{1+z'} dz'\right\rbrace }~.
\label{eq:hubble_recon}
\eer
In this case, if $\om$ is wrongly specified then, in a maximum likelyhood approach,
 the DE parameters $w_0,w_1$
will adjust to make $H(z)$ as close to its real
%(read `observationally
%determined')
value as possible, leading once more to an erroneous reconstruction of
the cosmic equation of state.

These two factors: the larger errors caused by the double differentiation
of a noisy quantity ($D_L$) and the strong dependence of $w(z)$ on an
uncertain quantity ($\om$) adversely impact the cosmological reconstruction
of the EOS making it difficult to differentiate a cosmological constant
from evolving DE from an analysis of $w(z)$ alone.

\subsection{The Om diagnostic introduced}

\begin{figure*}[ht]
\centerline{ \psfig{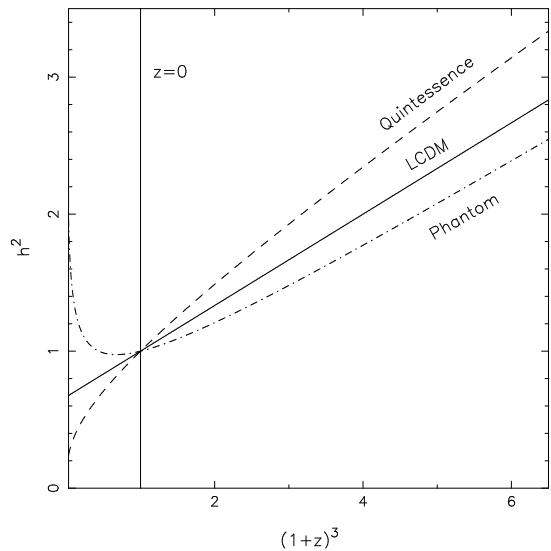} }
\bigskip
\caption{\small
The Hubble parameter squared is plotted against the cube
of $1+z$ for Quintessence ($w=-0.7$, dashed), LCDM ($w=-1$, solid) and Phantom
($w=-1.3$, dot-dash). The universe is assumed to be spatially flat and
$\Omega_{\rm DE} = 2/3$ in all models.
For LCDM the plot $h^2$ vs $(1+z)^3$ is a {\underline {straight line}} whereas for P and Q
this line is {\underline {curved}} in the interval $-1<z\lleq 1$.
This forms the basis for the observation that $Om(x_1,x_2) \equiv Om(x_1) - Om(x_2) = 0$ in LCDM,
while $Om(x_1,x_2) > 0$ in Quintessence and $Om(x_1,x_2) < 0$ in Phantom,
if $x_1 < x_2$. Thus $Om(x_1,x_2)$ furnishes us with a {\em null test}
for the cosmological constant.
(At $z<0$ the Hubble parameter for Phantom diverges at the
`Big Rip' future singularity, while for Quintessence $h(z) \to 0$ as $z \to -1$.
LCDM approaches the de Sitter space-time at late times.)
%The points marked $x_1,x_2$ are where the $Om$ diagnostic is determined.
%in relations
%(\ref{eq:lam}) - (\ref{eq:phantom}).
}
\label{fig:LCDM}
\end{figure*}

In this paper we suggest an alternative route which enables us to distinguish
LCDM from other DE models without directly involving the cosmic EOS.
Our starting point is the Hubble parameter which is used to determine
the $Om$ diagnostic
\beq
Om(x) \equiv \frac{h^2(x)-1}{x^3-1},~~ x=1+z~,~h(x) = H(x)/H_0~.
\label{eq:om}
\eeq
%where $h(x) = H(x)/H_0$.

\begin{figure*}[ht]
\centering
\begin{center}
$\begin{array}{@{\hspace{-1.0in}}c@{\hspace{0.0in}}c}
\multicolumn{1}{l}{\mbox{}} &
\multicolumn{1}{l}{\mbox{}} \\ [-0.20in]
\epsfxsize=3.2in
\epsffile{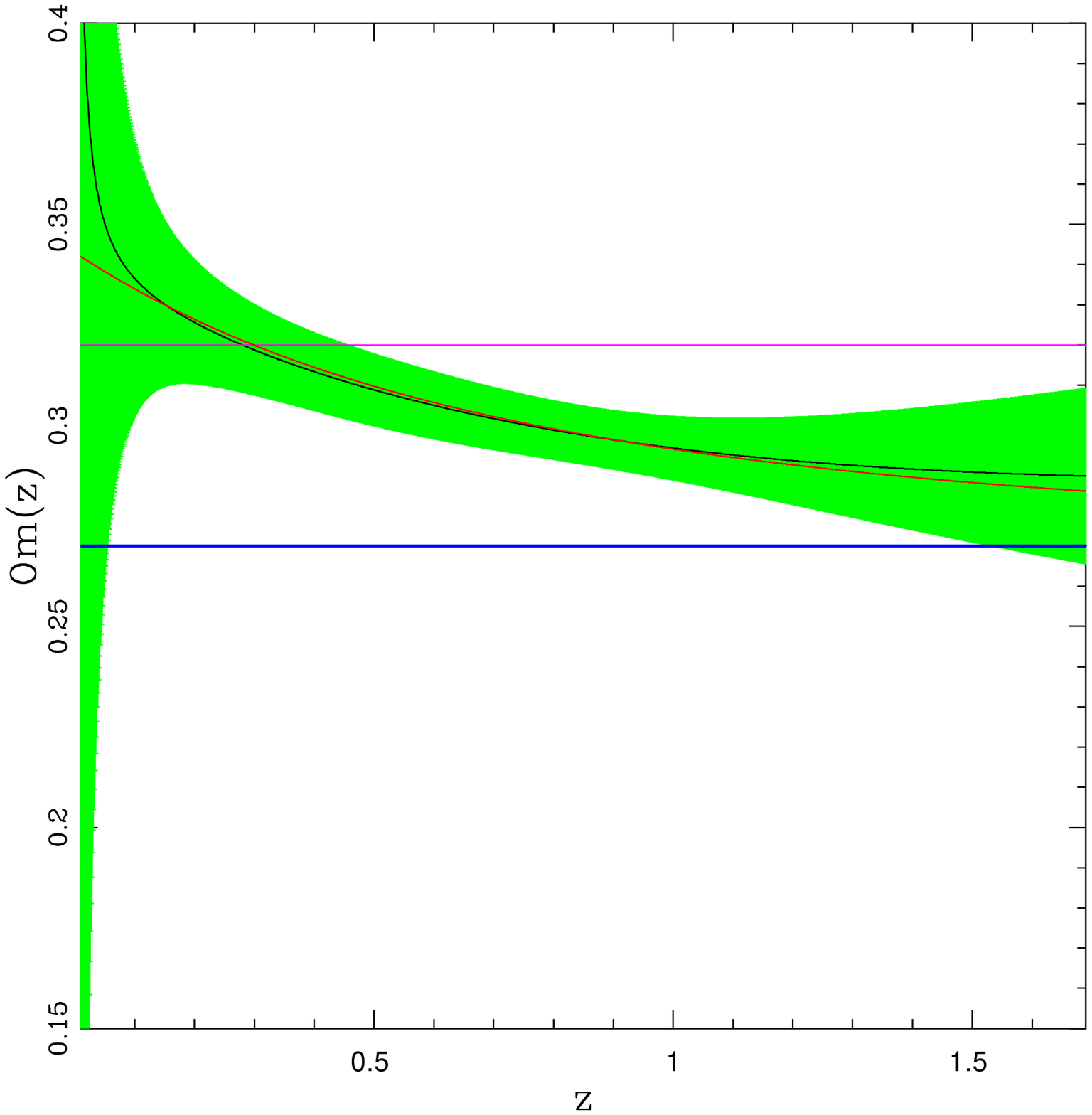} &
\epsfxsize=3.2in
\epsffile{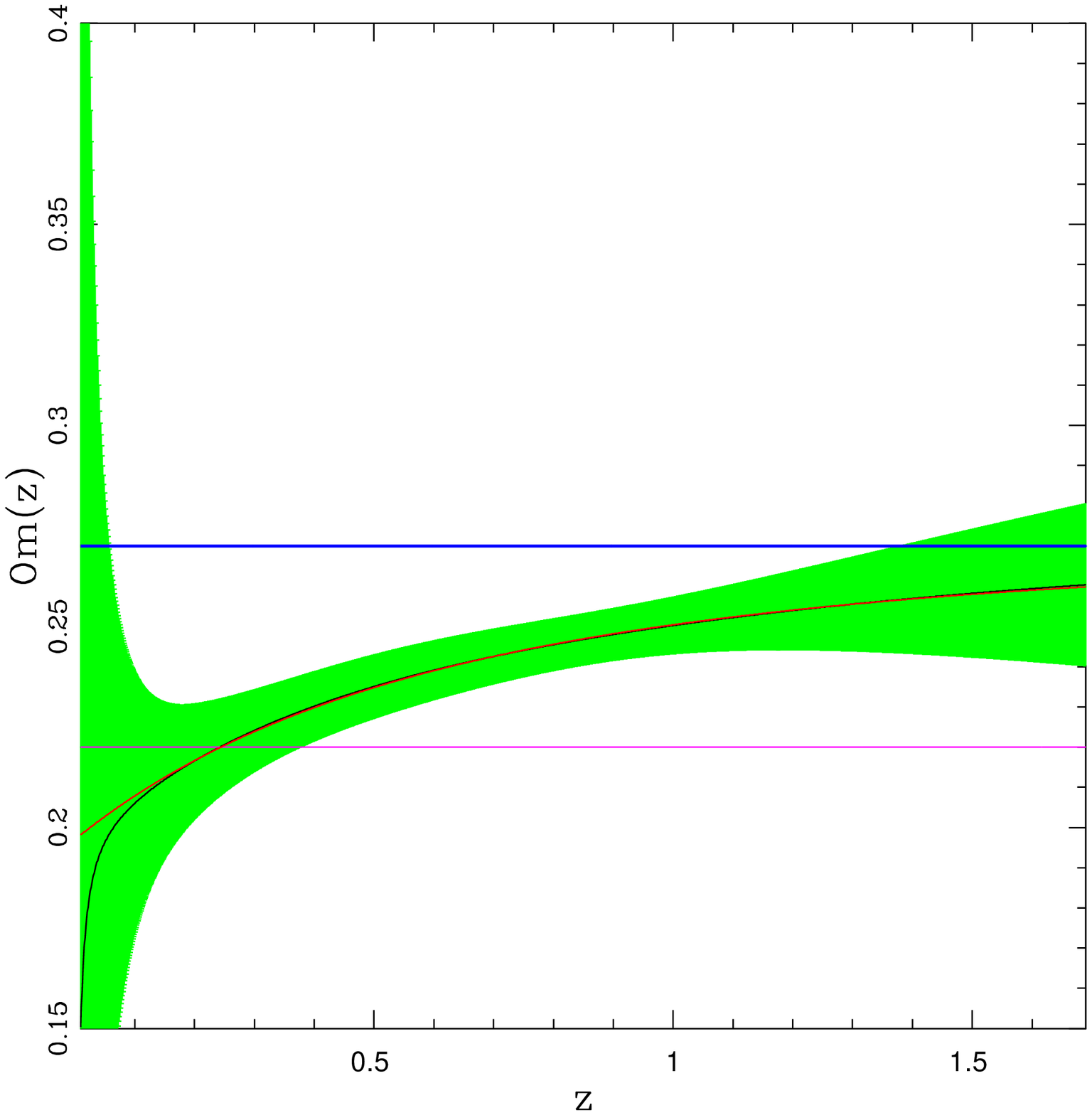} \\
\end{array}$
\end{center}
\vspace{0.0cm}
\caption{\small
The {\em left panel} shows the $Om(z)$ diagnostic reconstructed
for a fiducial {\em quintessence} model with
$w=-0.9$ and $\om = 0.27$ (black line, green shaded region shows $1\sigma$ CL,
the red line is the exact analytical result for $Om$).
The horizontal blue line shows the value of $Om$ for a
$\Lambda$CDM model with the same value of $\om$ as quintessence. Note that any
horizontal line in this figure represents
$\Lambda$CDM with a {\em different value of $\om$.}
For instance $\Lambda$CDM with $\om = 0.32$ is shown by the horizontal magenta line.
As this figure shows, the negative curvature of quintessence allows us to
distinguish this model from (zero-curvature) $\Lambda$CDM independently of
the current value of the matter density.
The {\em right panel} shows the $Om(z)$ diagnostic reconstructed for a fiducial
{\em phantom} model with
$w=-1.1$ and $\om = 0.27$ (black line, green shaded region shows $1\sigma$ CL).
%Again, the red line is the exact analytical result for $Om$ while the
%horizontal blue line shows $Om$ for a $\Lambda$CDM model with $\Omega_m = 0.27$.
The positive curvature of phantom allows us to
distinguish this model from (zero-curvature) $\Lambda$CDM independently of
the current value of the matter density.
For instance, phantom can easily be distinguished from $\Lambda$CDM both
with the correct $\om = 0.27$
(horizontal blue) as well as
incorrect $\om = 0.22$
(horizontal magenta).
(The non-parametric reconstruction scheme suggested in \cite{arman} has been employed on
SNAP quality data for this reconstruction.)
}
\label{fig:om1}
\end{figure*}

For dark energy with a constant equation of state $w=const$,
\beq
h^2(x) = \Omega_{0m}x^3 + (1-\Omega_{0m})x^\alpha, ~~\alpha = 3(1+w)
\label{eq:hubble}
\eeq
(we assume that the universe is spatially flat for simplicity).
Consequently,
\beq
Om(x) = \Omega_{0m} + (1-\Omega_{0m})\frac{x^\alpha - 1}{x^3-1}~,
\eeq
from where we find
\beq
Om(x) = \Omega_{0m}
\label{eq:om1}
\eeq
in LCDM, whereas $Om(x) > \Omega_{0m}$ in quintessence $(\alpha > 0)$ while
$Om(x) < \Omega_{0m}$ in phantom $(\alpha < 0)$.
We therefore conclude that:
{\underline{$Om(x) - \Omega_{0m} = 0$ iff DE is a cosmological
constant}} \footnote{Note that since $\Omega_\Lambda+\Omega_{0m} \simeq 1$
 in LCDM, this model contains
SCDM $(\Omega_{0m}=1, \Omega_\Lambda=0)$ as an important limiting case.
Consequently the $Om$ diagnostic cannot distinguish between
large and small values of the cosmological constant
unless the value of the matter density is independently known.
An identical degeneracy exists for the statefinder $r$ as
discussed in \cite{statefinder}.}.
 In other words, the $Om$ diagnostic provides us with a {\em null test}
of the cosmological constant.
This is a simple consequence of the fact that
$h^2(x)$ plotted against
$x^3$ results in a straight line for LCDM, whose slope is given by
$\Omega_{0m}$, as shown in figure \ref{fig:LCDM}.
%Note that the discriminatory power of $Om$ is based on the fact that
%$h^2$, when plotted against $x^3$, is a straight line {\em only for} LCDM.
For other DE models the line describing $Om(x)$ is curved, %as shown in figure \ref{fig:LCDM}.
since the equality
\beq
\frac{dh^2}{dx^3} = constant~,
\label{eq:slope}
\eeq
(which always holds for LCDM)
%is consistently obeyed only in (spatially flat) LCDM,
%while other DE models satisfy (\ref{eq:slope}) only
is satisfied in quintessence/phantom type models only
at redshifts significantly greater than unity,
when the effects of DE on the expansion rate can safely be
ignored. As a result the efficiency of the $Om$ diagnostic improves at low $z \lleq 2$
precisely where there is likely to be an abundance of cosmological data in the
coming years !
%On the other hand, for quintessence as well as phantom the line describing
%$Om(x)$ is curved, which helps
%distinguish these models from LCDM even if the value of the matter
%density is not accurately known -- see figure \ref{fig:om1}.

In practice, the
construction of $Om$ requires a knowledge of the Hubble parameter, $h(z)$, which can be
determined using a number of model independent approaches
\cite{statefinder,arman,wang_teg05}. In figure \ref{fig:om1} we show the
$Om$ diagnostic reconstructed from SNAP-quality data using the non-parametric
prescription of \cite{arman}. One clearly sees that
for quintessence as well as phantom the line describing
$Om(x)$ is curved, which helps
distinguish these models from LCDM even if the value of the matter
density is not accurately known.

\begin{figure*}[ht]
\centerline{ \psfig{figure=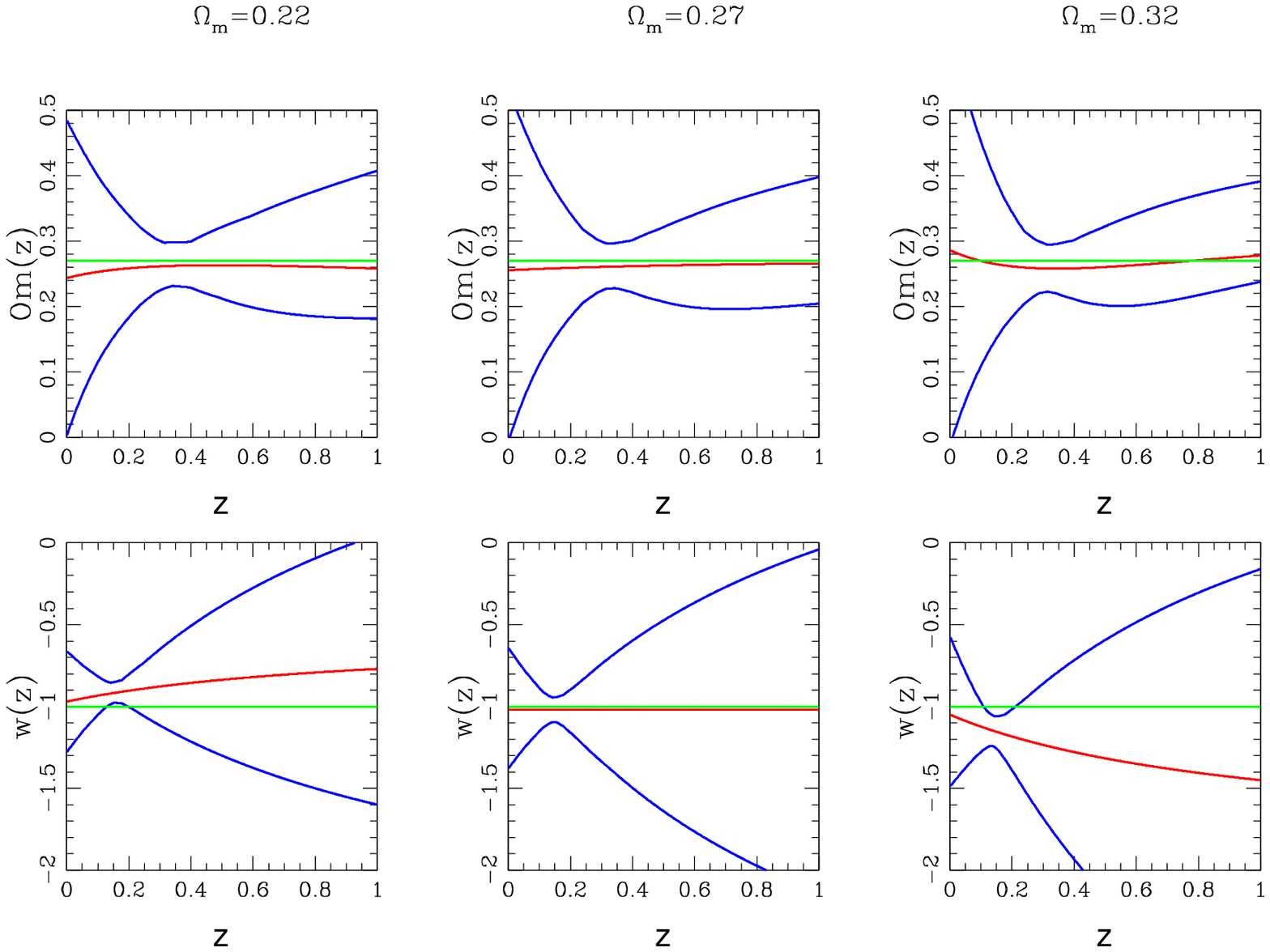,width=0.90\textwidth,angle=0,bbllx=40bp,bblly=172bp,bburx=590bp,bbury=600bp,clip=true} }
\bigskip
\caption{\small
Reconstructed $Om(z)$ and $w(z)$ from SNLS supernovae data using the CPL
ansatz (\ref{eq:cpl}) and
 assuming three different values $\Omega_m = 0.22, 0.27,
0.32$ for the matter density.
Notice that while the best fit value of $Om(z)$ is virtually independent
of the redshift (top panel, red curve) and is therefore consistent with 
LCDM (with $\Omega_m = 0.27$: green line), the reconstructed EOS strongly depends upon the value of
the matter density. Thus, for the same data set, the best fit value of $w(z)$ is suggestive of
quintessence for $\Omega_m = 0.22$, LCDM for $\Omega_m = 0.27$
and phantom for $\Omega_m = 0.32$, while $Om(z)$ favours LCDM throughout.
%On the other hand $Om(z)$ appears to
%be much more robust against variation in $\Omega_m$
%and is strongly suggestive of LCDM which is shown by the
%horizontal green line in the top panel for
%$Om (\equiv \Omega_m) =0.27$. %for LCDM which is consistent with
%the SNLS data.
Note that the small variations in $Om(z)$ in the three upper panels are a
consequence of the CPL ansatz which requires, as input, the value of the
matter density $\Omega_m$. A non-parametric ansatz such as \cite{arman},
or the parametric ansatz \cite{saini00},
would have led to a uniquely reconstructed $Om(z)$ with no dependence on
$\Omega_m$. Blue lines show $1\sigma$ error bars.
}
\label{fig:om_snls}
\end{figure*}

\begin{figure*}[ht]
\centerline{ \psfig{figure=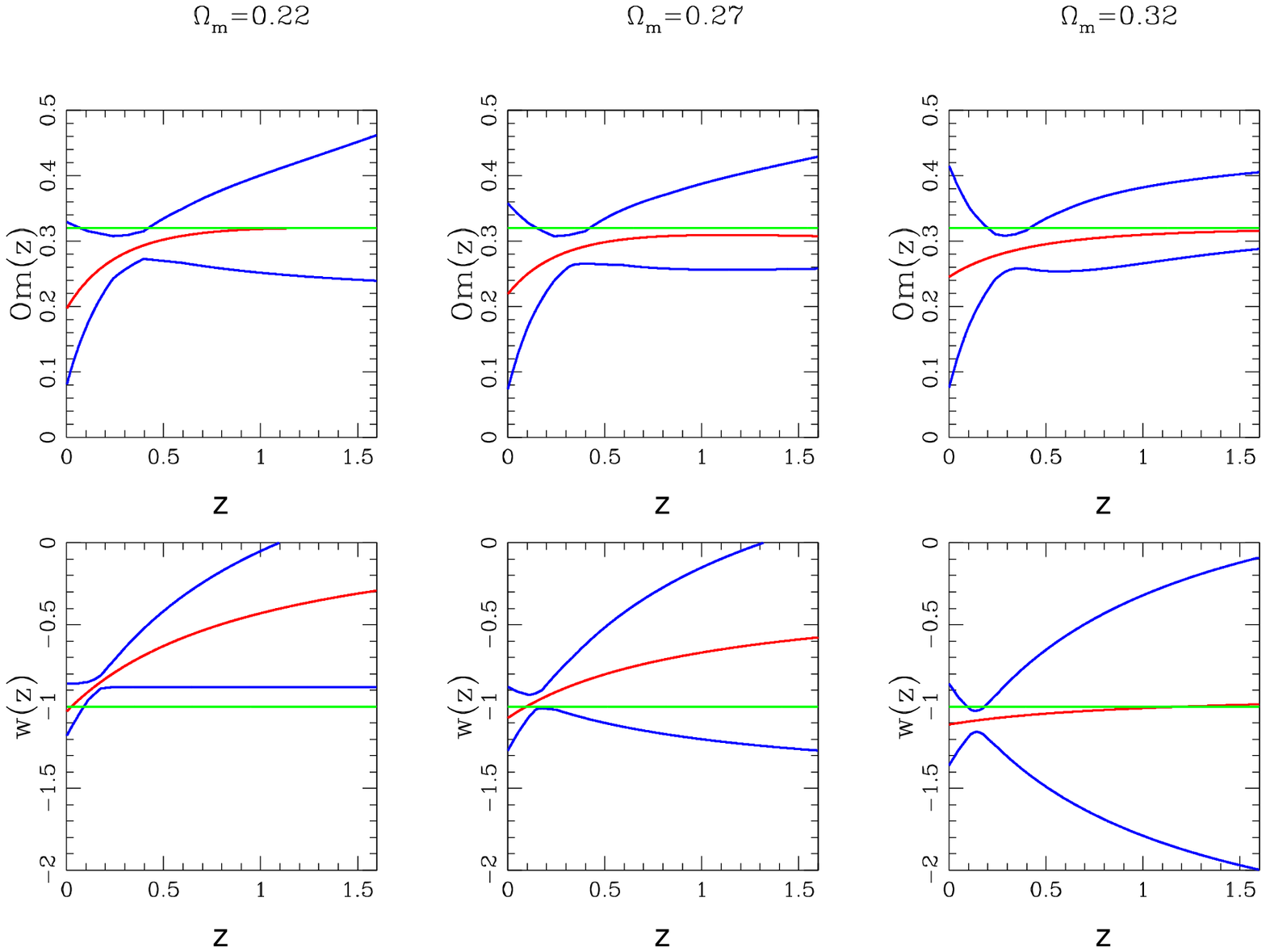,width=0.90\textwidth,angle=0,bbllx=40bp,bblly=172bp,bburx=590bp,bbury=600bp,clip=true} }
\bigskip
\caption{\small
Reconstructed $Om(z)$ and $w(z)$ from recent Union supernovae data using the CPL ansatz
(\ref{eq:cpl}) and assuming three different values $\Omega_m = 0.22, 0.27, 0.32$ for the matter density. $Om(z)$ appears to be much more robust against variation in $\Omega_m$ in comparison with $w(z)$. The horizontal green line in the top panel indicates value
of $Om (\equiv \Omega_m)= 0.32$ for LCDM model. The blue lines show $1\sigma$ error bars. Though LCDM is still consistent with the Union data, this consistency is not quite as strong as
it was for the SNLS data shown in the previous figure. The top panel clearly indicates that
 evolving DE is also perfectly consistent with Union data.
}
\label{fig:om_union}
\end{figure*}

Clearly, a comparison of $Om$ at two different
redshifts can lead to insights about the nature of DE {\em even if
the value of $\Omega_{0m}$ is not accurately known}. Thus, the
two-point difference diagnostic
\beq
Om(x_1,x_2) \equiv Om(x_1) - Om(x_2) =
%\frac{h^2(x_1)-1}{x_1^3-1} - \frac{h^2(x_2)-1}{x_2^3-1}~,
(1-\Omega_{0m})\left\lbrack\frac{x_1^\alpha - 1}{x_1^3-1}-\frac{x_2^\alpha - 1}{x_2^3-1}
\right\rbrack~,
\label{eq:om12}
\eeq
can serve as a {\em null} test of the cosmological constant hypothesis, since
\beq
Om(x_1) = Om(x_2) ~ ~~~ (\Lambda-term)~.
\label{eq:lam}
\eeq
In other words, {\underline{$Om(x_1,x_2) = 0$ iff DE is a cosmological constant;}}
$Om(x_1,x_2) > 0$ for quintessence while
$Om(x_1,x_2) < 0$ for phantom ($x_1 < x_2$).
Thus, the value of $Om$ determined at two redshifts can help distinguish between DE models
{\em without reference either to the matter density or $H_0$} !

We have reconstructed the $Om$ diagnostic and the cosmic EOS for
two SNe data sets: SNLS \cite{snls} and Union \cite{union}.

The SNLS (Supernova Legacy Survey) dataset contains 115 Type Ia supernovae (SNe Ia) in the
range $0.1 < z < 1.0. $
The Union dataset \cite{union} is a new compilation of SNe Ia and consists
of 307 SNe after selection cuts, includes the recent samples from the SNLS \cite{snls}
and ESSENCE Surveys \cite{essence}, older datasets, as well as the recently extended
dataset of distant supernovae observed with HST \cite{HST}.

 Results for the SNLS dataset, shown
 in figure \ref{fig:om_snls}, indicate that while the EOS is quite sensitive
to the value of the matter density, the $Om$ diagnostic is not.
Note that the three distinct models of dark energy in figure \ref{fig:om_snls}
result in virtually the same luminosity distance since:
(i) $\chi^2 = 110.93$ for $\om=0.32$ (Phantom),
(ii) $\chi^2 = 110.99$ for $\om=0.28$ (LCDM),
(iii) $\chi^2 = 111.02$ for $\om=0.22$ (Quintessence).
This shows that different values of $\om$ and $w(z)$ can provide an
excellent fit to the same set of data, as originally pointed out by
\cite{maor02}. From the upper panel of figure \ref{fig:om_snls}
 we find that the
$Om$ diagnostic is
 virtually independent of the input value of
$\om$ and its flat form
 is suggestive of LCDM \footnote{Note that although $Om(z) \simeq constant$,
its value in the
upper left and right panels of figure \ref{fig:om_snls} is not equal to $\om$. Therefore from (\ref{eq:om1})
it follows that the best fit DE model in these two cases is not LCDM.
We conclude that although the behavior of Om(z) appears tantalizingly similar
 to what one
would expect for the cosmological constant, the severe
degeneracies between $\om$ and $w$ prevent us from drawing a firm conclusion about the
nature of dark energy on the basis of SNLS data alone.}.
Note that the degeneracy between evolving dark energy and
the cosmological
constant can be broken by studying the behavior of $Om(z)$ at higher redshifts.
For most models the effect of DE on expansion becomes negligible
for $z > {\rm few}$ leading to
$Om(z) \to \om$ at moderately high redshifts. Consequently the differential diagnostic
$Om(z_1,z_2)$ evaluated at $|z_1-z_2| \gg 1$ can help break the degeneracy caused by
uncertainties in the value of $\om$. For the parameters in figure \ref{fig:om_snls},
$Om(z_1,z_2)=0$ for the cosmological constant, while
$|Om(z_1,z_2)|=0.05$ for evolving DE; see section \ref{sec:metamorph} for a related discussion. 
%The $Om$ diagnostic is unaffected by this degeneracy
%(between $\om$ and $w$) since its form
% appears to be virtually independent of the input value of
%$\om$ (and is suggestive of LCDM) as shown in the upper panel of figure \ref{fig:om_snls}.
%This leads us to conclude that $Om$ is a robust indicator of DE.

Figure \ref{fig:om_union} shows results for the more recent Union dataset.
Again we see that the behaviour of the EOS can range from being quintessence-like
(for $\om=0.22$) to being phantom-like (for $\om=0.32$).
The behaviour of $Om$ is less sensitive to the value of the matter density and leads us
to conclude that while
a cosmological constant appears to be strongly preferred
by SNLS, constraints from the Union dataset allow
evolving DE as well as $\Lambda$.

One can improve the efficiency of the $Om$ diagnostic (\ref{eq:om12})
by determining it selectively in regions %of redshift space
where there is better quality data.
The error in the reconstructed value of
the Hubble parameter is \cite{tegmark02}
\beq
\label{eq:tegmark}
\frac{\delta H}{H}(z)\propto \frac{\sigma}{N(z)^{1/2}}~,
\eeq
where $N(z)$ is the number of supernovae in a given redshift interval
and $\sigma$ is
the noise of the data. Since $N(z)$ is never likely to be a perfectly
uniform distribution, there will always be regions where $N(z)$ is larger
and $H(z)$ better reconstructed. Consequently, by determining
$Om(z_1,z_2)$ selectively in such regions, one can improve the efficiency
of this diagnostic by `tuning it' to the data. %as illustrated in figure....

\subsection{Dark Energy Metamorphosis}
\label{sec:metamorph}

\begin{figure*}
\centerline{ \psfig{figure=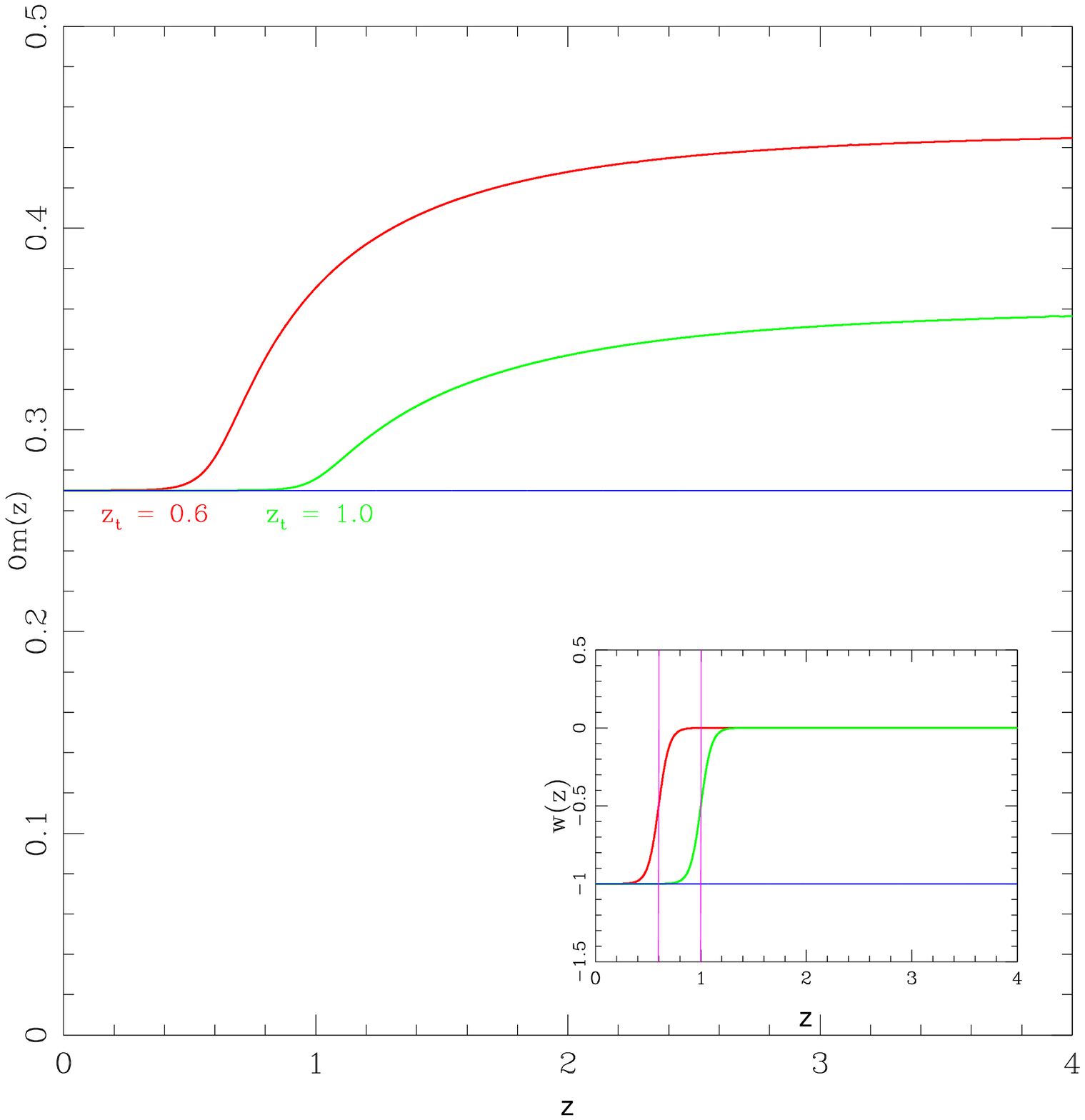,width=0.55\textwidth,angle=0} }
\bigskip
\caption{\small The $Om$ diagnostic is shown for two
tracker models which mimick LCDM at low redshift ($z < z_t$) and dark matter at
high redshift ($z > z_t$).
The horizontal blue line shows LCDM.
The inset shows the EOS for the tracker's as a function of redshift.
}
\label{fig:track}
\end{figure*}

An important example of quintessence is provided by
tracker DE models, which give rise to cosmic acceleration at late times
while earlier, during the radiation and matter dominated epochs, the density in the
tracker remains proportional to the background matter density \cite{bcn00,sw00}.
This last property leads to
$\rho_{\rm track}/\rho_{\rm B}
\simeq constant \ll 1$ at $z > z_t$, where $\rho_{\rm B}$ is the background
density of matter or radiation and $z_t$ is the redshift when tracking ends.
As an example consider the double exponential model \cite{bcn00}
$V(\phi) = M^4\lbrack\exp{(-\alpha\phi)} + \exp{(-\beta\phi})\rbrack$ with
$\alpha \gg \beta$, $\beta \ll 1$, which has the attractor solution
$\Omega_\phi = 3(1+w_B)/\alpha^2$,
$w_\phi = w_B$, at high redshift $z > z_t$,
while $w_\phi \simeq -1 + \alpha^2/3$ at the present
time. Since these models behave like quintessence at late times, their behavior
is similar to that shown in figure 1 for a typical quintessence model.
Consequently these models may be distinguished from LCDM by applying the
$Om$ diagnostic shown in figure 1.
An interesting limiting case corresponds to {\em metamorphosis} models which have $w_0 \simeq -1$
today and $w \to 0$ at earlier times
\cite{metamorphosis,george}.
%A DE model which
%behaves LCDM at present and is dust-like at earlier times satisfies:
For such models $Om(x) = \Omega_{0m}$ for $x < x_t$ and $Om(x) = \tilde{\Omega}_{0m}$,
for $x \gg x_t$ where
\beq
%\Omega_{0m} < \tilde{\Omega}_{0m} = \frac{\rho_m + \rho_{\rm track}}{\frac{3H_0^2(1+z)^3}{8\pi G}}~.
\tilde{\Omega}_{0m} \simeq \Omega_{0m} + \frac{1-\Omega_{0m}}{(1+z_t)^3}~.
\eeq
Consequently, the $Om$ diagnostic applied to data at low and high redshift, may help
distinguish between tracker DE and LCDM as shown in figure \ref{fig:track}.
Tracker behavior can also arise in modified gravity theories such as
Braneworld models \cite{brane} and scalar-tensor cosmology \cite{beps00}.
The growth of density perturbations provides a complementary means of distinguishing
these models from LCDM.

An important property of the $Om$ diagnostic is that the value of the
cosmological density parameter $\om$ does not enter into its
definition (\ref{eq:om}) explicitly. As a result the diagnostic
relation $Om(x_1,x_2) = 0$ (LCDM) does
not require an a-priori knowledge of the matter density and therefore provide a means of
differentiating the cosmological constant from evolving DE models
even if uncertainties exist in the value of $\om$.
(Current observations suggest an uncertainty of at least 25$\%$ in the value of $\om$
\cite{wmap05}.)
For a constant EOS the dependence of $Om$ on the value of the matter density can be altogether
eliminated by constructing the ratio
\beq
{\cal R} = \frac{Om(x_1,x_2)}{Om(x_3,x_4)} \equiv
\frac{\left\lbrack\frac{x_1^\alpha - 1}{x_1^3-1}-\frac{x_2^\alpha - 1}{x_2^3-1}
\right\rbrack}{\left\lbrack\frac{x_3^\alpha - 1}{x_3^3-1}-\frac{x_4^\alpha - 1}{x_4^3-1}
\right\rbrack}~,
\label{eq:R}
\eeq
from where we see that the EOS encoded in the parameter $\alpha = 3(1+w)$ can
be determined from ${\cal R}$ without any reference whatsoever to the value of $\om$ !

\subsection{Influence of spatial curvature on $Om$}

The preceding analysis, which showed how the $Om$ diagnostic could
distinguish between alternative models of DE, was based on the
assumption that the universe was spatially flat. While
this may well be true with sufficient accuracy, especially within
the framework of the inflationary scenario which predicts
$|\Omega_k-1|\sim 10^{-5}$ today (due to the $l=0$ mode of
primordial scalar fluctuations), let us now consider the case of
small, but non-zero spatial curvature. Then $Om(x)$ acquires the
correction
\begin{equation}
\delta Om = \Omega_k \frac{x+1}{x^2+x+1} ~, ~~x=1+z~,
\end{equation}
effectively, which has a fixed functional dependence on $z$. So,
$|\delta Om|\le 2|\Omega_k|/3$ and decreases with the growth of $z$.
For the interval $-0.0175 < \Omega_k < 0.0085$ at the $95\%$ CL
admitted by the WMAP5 results \cite{wmap05}, this correction does
not exceed $0.01$ (or $\sim 4\%$ of the value of the $Om$ itself).

In this case (\ref{eq:om12}) is modified to 
\beq Om(x_1,x_2) =
\Omega_{\rm DE}\bigg\lbrack\frac{x_1^\alpha-1}{x_1^3-1}
-\frac{x_2^\alpha-1}{x_2^3-1}\bigg\rbrack\
+ \Omega_k\bigg\lbrack\frac{x_1^2-1}{x_1^3-1}
-\frac{x_2^2-1}{x_2^3-1}\bigg\rbrack~, 
\label{eq:om_curv}
\eeq
and we assume $x_2>x_1$ as in figure  1.

The influence of the curvature term can be estimated by a simple `back of the envelope'
calculation which we carry out for
quintessence (Q) and phantom (P). We assume $\Omega_{\rm DE} = 0.7$,
$\Omega_k = -0.0175$, $z_1=0.1, z_2=1$
and $w=-0.9$ for Q while $w=-1.1$ for P.
In the case of Quintessence we find $Om(x_1,x_2) = 0.042$ when the curvature
term is included in (\ref{eq:om_curv}) and $Om(x_1,x_2) = 0.038$ when it is
excluded. Thus the inequality $Om(x_1,x_2) > 0$, which generically holds for
quintessence models,
appears to be quite robust, since the contribution
(read `contamination') from the curvature term is only a fraction ($9\%$)
of the `signal' from DE.
Similar results are obtained for Phantom: $Om(x_1,x_2) = -0.037$ when the curvature
term is included and $Om(x_1,x_2) = -0.041$ when it is not.
The presence of curvature leads, once more, to a $9\%$ change in our estimation
of $Om$ leading us to conclude that the phantom inequality $Om(x_1,x_2) < 0$ is robust.
(Of course, as $w_{\rm DE} \to -1$, the relative influence of curvature in
(\ref{eq:om_curv}) becomes significant and can dominate the `signal' from DE
models which are very close to LCDM; see also \cite{curvature}.)

\section{The Acceleration probe}

In the previous section we showed how the difference between the value of the
Hubble parameter at nearby redshifts could be used to construct a
null diagnostic for the LCDM model.
In this section we construct another
dimensionless quantity which could prove useful for determining the onset of
cosmic acceleration in DE models which has also been the focus of other recent
studies \cite{waga,lima}.

Our diagnostic, {\em acceleration probe}, is the
mean value of the deceleration parameter
\beq
{\bar q} = \frac{1}{t_1-t_2}\int_{t_2}^{t_1} q(t) dt~.
\eeq
Since
\beq
q(t) = \frac{d}{dt}\left (\frac{1}{H}\right ) - 1~,
\eeq
it follows that {\em acceleration probe} can be written in the following
simple form
\beq
1+{\bar q} = \frac{1}{\Delta t}\left (\frac{1}{H_1} - \frac{1}{H_2}\right )
\label{eq:q1}
\eeq
where
$\Delta t = t_1-t_2 \equiv (t_0-t_2) - (t_0-t_1)$, and
\beq
t_0-t(z) = \int_0^z\frac{dz}{(1+z)H(z)}
\label{eq:q2}
\eeq
is the cosmic look-back time (also see \cite{age,lima}).

Equation (\ref{eq:q1}) expresses the mean deceleration parameter
in terms of the look-back time and the value of the Hubble parameter at two
distinct redshifts.
%both of which can be reconstructed to good accuracy
%(see for instance \cite{arman,lima}). Since the look-back time was reconstructed to an %accuracy of
%better than $0.2\%$ while $H(z)$ was reconstructed to an accuracy $< 2\%$
%(at $z < 1$) using our smoothing approach,
%it follows that the {\tt q-probe} may also be reconstructed to
%very good accuracy, especially
%at low redshifts.
 From expressions (\ref{eq:q1}) and (\ref{eq:q2}) we find that, like the $Om$ diagnostic,
the acceleration probe ${\bar q}$ does not depend
upon the value of $\Omega_m$ and is, therefore, robust to uncertainties in the value of the
matter density.

\begin{figure*}[t]
\centerline{ \psfig{figure=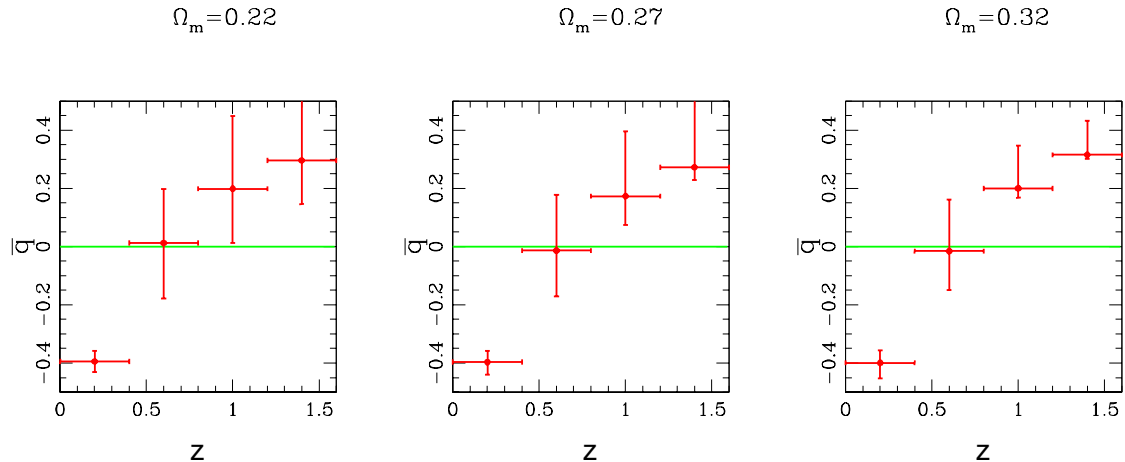,width=0.90\textwidth,angle=0,bbllx=40bp,bblly=340bp,bburx=590bp,bbury=600bp,clip=true} }
\bigskip
\caption{\small The diagnostic $\bar q$ is plotted in 4 bins using the recent Union supernovae
data. The CPL ansatz has been used for three different values of the matter density.
Error-bars in y-axis show $1-\sigma$ CL. Note that the value of the acceleration redshift
$0.4 \leq z_a \leq 0.8$ appears to be robust.}
\label{fig:qbar}
\end{figure*}

In figure \ref{fig:qbar} we show $\bar q$ obtained using Union supernovae and the CPL ansatz.
 The behaviour of $\bar q$ suggests $ 0.4 \leq z_a \leq 0.8 $ for the redshift at which the
universe began to accelerate. This result is independent of the value of the matter density.
Close to the acceleration redshift,
${\bar q} \simeq 0$, and one obtains
a very simple relationship linking the look-back time with the
Hubble parameter
\beq
\Delta t = \frac{1}{H_1} - \frac{1}{H_2}~,
\label{eq:q3}
\eeq
where $H_1$ and $H_2$ lie on `either side' of the acceleration redshift $z_a$ when $q(z_a) = 0$.
Since both the look-back time and the Hubble parameter can be reconstructed quite accurately (see for instance \cite{ss06,age}), it follows that one might be able to obtain the redshift of the acceleration epoch in a model independent manner using (\ref{eq:q3}).

%It is worth mentioning that $\bar q$ can also be obtained from discrete cosmological observations.
It is worth noting that the value of $\bar q$ can also be obtained from an accurate
determination of galactic ages.
%For instance measurements of ages of the relaxed galaxies can be used for this purpose.
In this case we do not require a continuous form of $H(z)$ to compute the look-back time.
We simply subtract the galactic ages at two distinct redshifts bins to determine $\Delta t$.
The same information can also be used to derive $H(z)$ \cite{age} since
\beq
H(z) = -\frac{1}{1+z} \frac{dz}{dt},
\label{eq:age}
\eeq
which is then used to determine $\bar q$.
However at present errorbars in observed galactic ages are large and the number of data is
 small, so it is unlikely that this method will be useful for determining $\bar q$ at this stage. In the
 nearby future, with better quality and quantity of data, $\bar q$ can be used as a model independent probe of the acceleration of the universe using distinct and uncorrelated cosmological data.

Note that the acceleration epoch is quite sensitive to the underlying DE model.
For DE with a constant equation of state
\beq
%(1+z_a)^{-3w} = -(1+3w)\frac{\Omega_{DE}}{\om}~,
1+z_a = \left (\frac{|1+3w|\Omega_{0DE}}{\om}\right )^\frac{1}{|3w|}~,
\eeq
and so an accurate determination of $z_a$ using (\ref{eq:q3}) could provide useful insights into the nature of DE.

\section{Determining $Om$ and the {\em acceleration probe} from SNe, BAO and CMB}

\begin{figure*}[t]
\centerline{ \psfig{figure=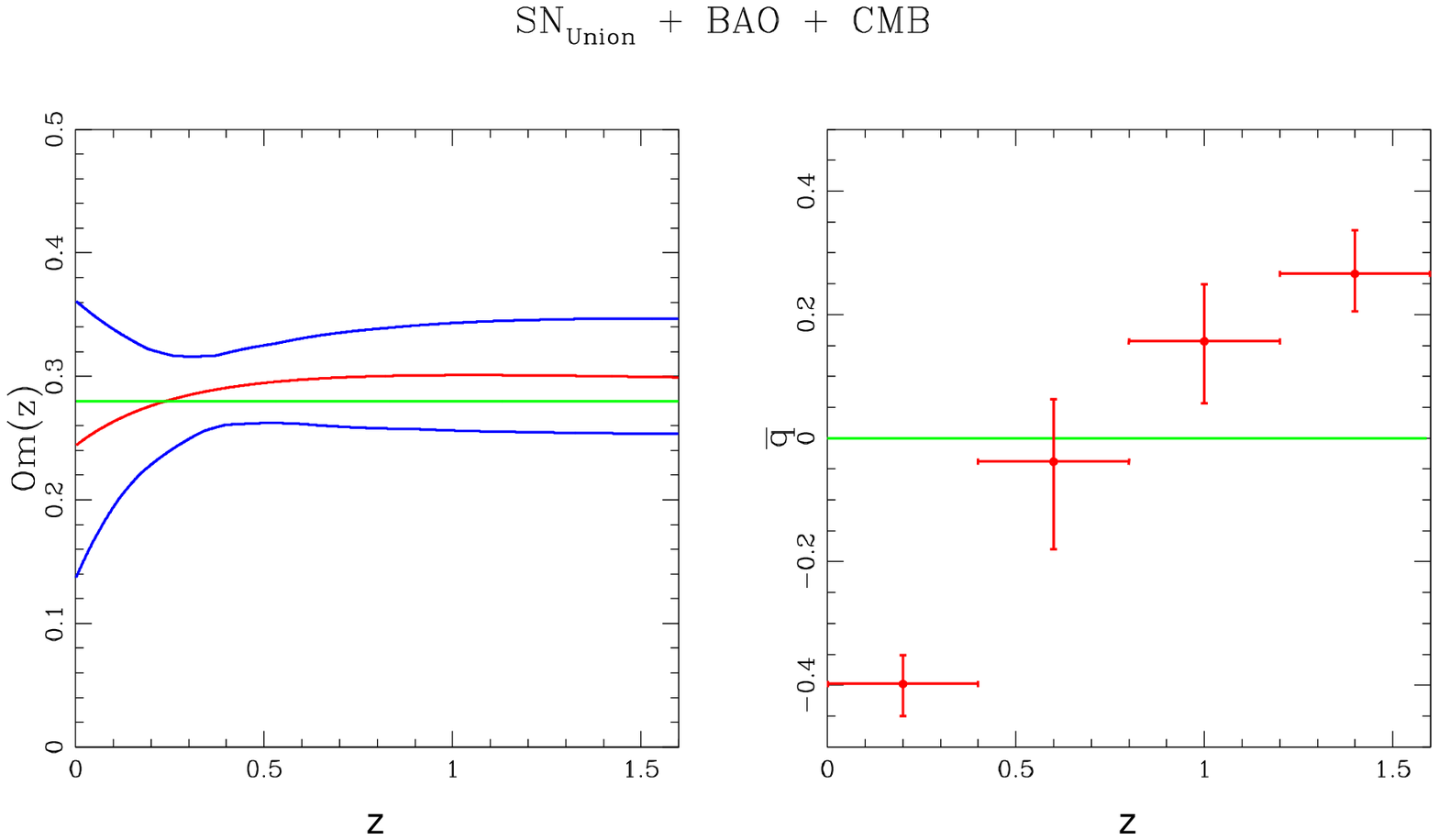,width=0.90\textwidth,angle=0,bbllx=30bp,bblly=150bp,bburx=590bp,bbury=500bp,clip=true} }
\bigskip
\caption{\small Two new diagnostics, $Om(z)$ (left panel) and $\bar q$ (right panel) are plotted using
a combination of supernovae, BAO and CMB data. The CPL ansatz has been used assuming the
 matter density also to be a free parameter. Blue lines in left panel and red crosses in right panel show $1\sigma$ errorbars.}
\label{fig:all}
\end{figure*}

\begin{figure*}[t]
\centerline{ \psfig{figure=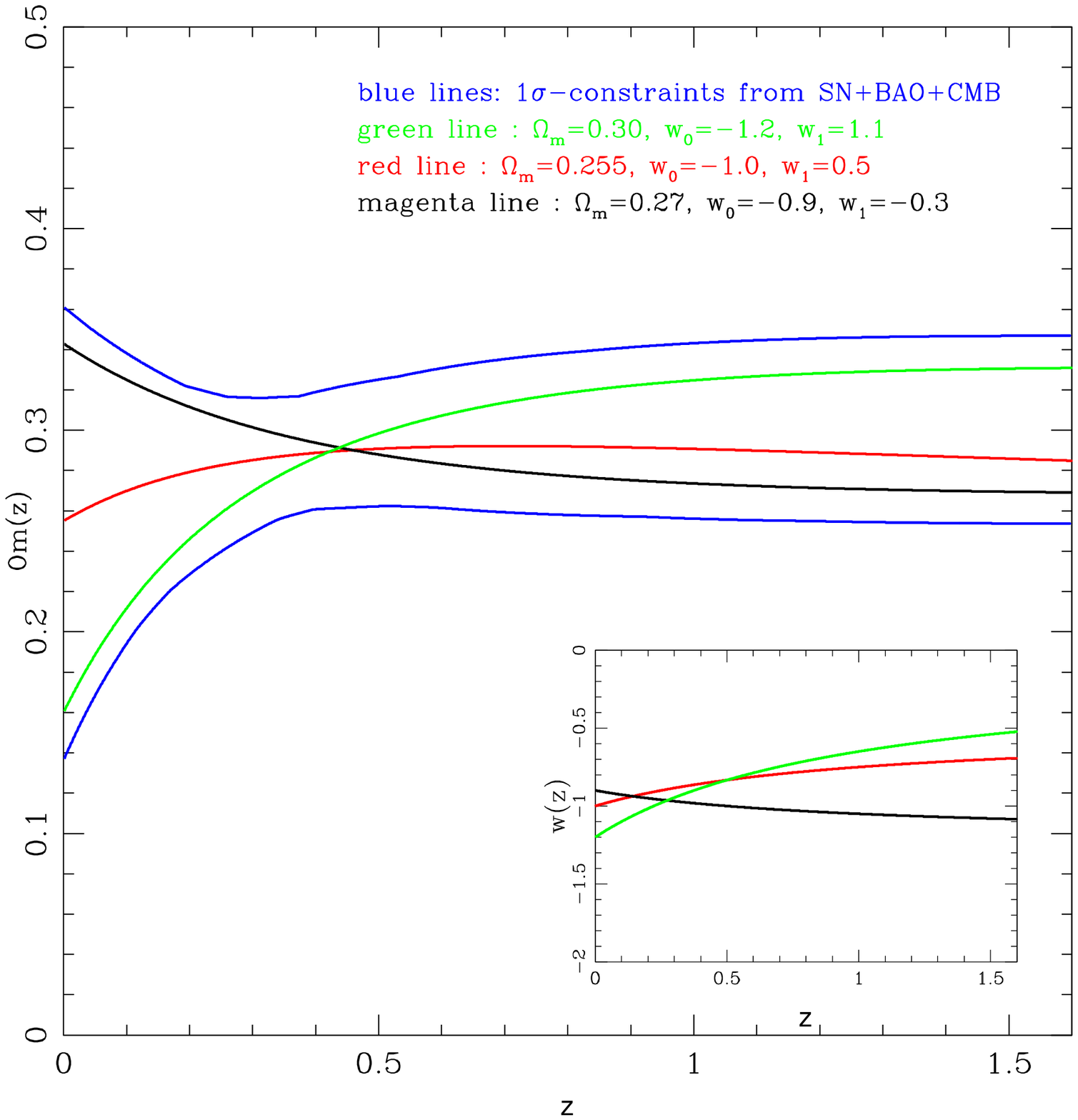,width=0.50\textwidth,angle=0,bbllx=30bp,bblly=150bp,bburx=590bp,bbury=700bp,clip=true}}
\bigskip
\caption{\small The blue lines in the main figure show $1 \sigma$
constraints on $Om(z)$ from a combination of SNe, BAO and CMB data. Also shown are
values for $Om(z)$ from three DE models all of which
are consistent with the data at the $1 \sigma$ level. The green line is DE
 with $\Omega_m=0.3, w_0=-1.2, w_1=1.1$, this model crosses the phantom divide at
$w=-1$. The red line shows a metamorphosis model with $\Omega_m=0.255, w_0=-1.0, w_1=0.5$,
while magenta shows the model with $\Omega_m=0.27, w_0=-0.9, w_1=-0.3$.
In all cases the CPL ansatz (\ref{eq:cpl}) has been used and
the bottom-right corner of the figure shows the EOS for these diverse DE models.
Note that in both the main figure as well as the inset, LCDM corresponds to a horizontal
straight line (not shown).}

\label{fig:models}
\end{figure*}

In this section we determine our two new diagnostics, $Om$ and $\bar q$, from a
 combination of: (i) the Union supernovae data set \cite{union}, (ii) data from
baryon acoustic oscillations (BAO) \cite{bao}, (iii) WMAP5 CMB data \cite{wmap5}.

Acoustic oscillations in the photon-baryon plasma prior to recombination give rise to
a peak in the correlation function of galaxies. This effect has recently been measured
in a sample of luminous red galaxies observed by the Sloan
Digital Sky Survey and leads to the value
\cite{bao}
\beq
A = \frac{\sqrt{\om}}{h(z_1)^{1/3}}~\bigg\lbrack ~\frac{1}{z_1}~\int_0^{z_1}\frac{dz}{h(z)}
~\bigg\rbrack^{2/3} = ~0.469 \left(\frac{n}{0.98}\right)^{-0.35} \pm 0.017~,
\eeq
where $h(z) = H(z)/H_0$
and $z_1 = 0.35$ is the redshift at which the acoustic scale has
been measured.
The 5 year Wilkinson Microwave Anisotropy Probe (WMAP5) results,
when combined with the results from BAO
yield $n=0.961$ for the spectral index of the primordial power spectrum \cite{wmap5,LAMBDA}.

We also use the following value for the CMB `shift parameter'
(the reduced distance to the last scattering surface) deduced from WMAP5
\beq
R = \sqrt{\om}\int_0^{z_{\rm ls}} \frac{dz}{h(z)} = 1.715 \pm 0.021~,
\eeq
where $z_{\rm ls} = 1089$.
We use the two constraints, $A$, $R$ together
with the Union SNe data set to determine $Om(z)$ and $\bar q$.
The CPL ansatz (\ref{eq:cpl})
has been used to parametrize the expansion history, and all three parameters in this
ansatz: $\om$, $w_0$ and $w_1$, are treated as being free in our maximum likelyhood
routine. Our results are summarized in figure \ref{fig:all}.
We find that LCDM is in excellent agreement with the data but other DE models fit the data
too. These include quintessence and phantom models, some of which are
shown in
figure \ref{fig:models}.

\section{Conclusions}

In this paper we propose two new diagnostics for determining the
properties of dark energy. The first of these, $Om(z)$, is
constructed from the Hubble parameter and results in the identity
$Om(z) = \om$ for LCDM. For other DE models $Om(z)$ is a function of
the redshift. This allows one to construct a simple {\em null test}
to distinguish the cosmological constant from evolving DE. We find
that $Om$ is a robust diagnostic whose value can be determined
reasonably well even with current data. Unlike the
equation of state $w$, $Om$ relies only on a knowledge of the Hubble
parameter and depends neither on $H'(z)$ nor on quantities
like the growth factor $\delta(z)$. Errors in the reconstruction of
$Om$ are therefore bound to be smaller than those appearing in the
EOS. To diminish systematic uncertainties, different tests which 
allow for the possibility of reconstructing $H(z)$ from observational data
need to be used. $Om$ can be determined using parametric as well as
non-parametric reconstruction methods. It can shed light on the
nature of dark energy even if the redshift distribution of
supernovae is not uniform and the value of the dark matter density
is not accurately known. The second diagnostic, {\em acceleration
probe} ${\bar q}$, is the mean value of the deceleration parameter
over a small redshift interval. The {\em acceleration probe} depends
upon the value of the Hubble parameter and the look-back time and can be used
to determine the epoch when the universe began to accelerate.

We use the current type Ia SNe data in conjunction with BAO and CMB data to
estimate the values of $Om(z)$ and ${\bar q}$. Our results are
consistent with LCDM but do not exclude evolving DE models including
phantom and quintessence.

{\em Note added in proof:}
At $z \ll 1$, 
\beq
\frac{Om - \Omega_{0m}}{1- \Omega_{0m}} \simeq 1+w_0~, \nonumber\\
\eeq
which
suggests that the $Om$ diagnostic can be used to probe $w_0$ from data
available at low $z$, if $\Omega_{0m}$ is accurately known.

\section*{Acknowledgments}

VS acknowledges a stimulating conversation with J.A.S. Lima.
AAS acknowledges IUCAA hospitality as a visiting professor. He was also
partially supported by the grant RFBR 08-02-00923 and by the Scientific
Programme "Astronomy" of the Russian Academy of Sciences.

\end{document}